\documentclass[12pt]{article}
\usepackage[sort&compress,square,comma,numbers]{natbib}
\usepackage{amsmath,amssymb,graphicx,slashed}
\pdfoutput=1

\usepackage[
	colorlinks=true,
	citecolor=black,
	linkcolor=black,
	urlcolor=blue,
	hypertexnames=false]{hyperref}
	
\newcommand{\arXiv}[2]{\href{http://arxiv.org/pdf/#1}{{\tt #2/#1}}}
\newcommand{\arXivold}[1]{\href{http://arxiv.org/pdf/#1}{{\tt #1}}}

\newcommand{\beq}{\begin{eqnarray}}
\newcommand{\eeq}{\end{eqnarray}}

\def\tilde#1{\widetilde{#1}}
\DeclareMathOperator{\Tr}{Tr}

\renewcommand{\Re}{\mathop{\rm Re}}
\DeclareMathOperator{\sgn}{sgn}
\def\epsilon{\varepsilon}
\def\s{\slashed}

\begin{document}
\begin{center} 
{\huge \bf $S$-Duality and \\ \vspace*{0.25cm}
Helicity Amplitudes \vspace*{0.5cm}} 
\end{center}

\begin{center} 
{\bf \  Kitran Colwell and John Terning} \\
\end{center}
\vskip 8pt
\begin{center} {\it Department of Physics, University of California, Davis, CA 95616} 

\vspace*{0.1cm}

{\tt 
 \href{mailto:colwell@ms.physics.ucdavis.edu}{colwell@ms.physics.ucdavis.edu}, 
 \href{mailto:jterning@gmail.com}{jterning@gmail.com}}

\end{center}

\centerline{\large\bf Abstract}
\begin{quote}
We examine interacting Abelian theories at low energies and show that holomorphically normalized photon helicity amplitudes  transform into dual amplitudes under $SL(2,\mathbb{Z})$ as modular forms with weights that depend on the number of positive and negative helicity photons and on the number of internal photon lines. Moreover, canonically normalized helicity amplitudes transform by a phase, so that even though the amplitudes are not duality invariant,
their squares are duality invariant. We explicitly verify the duality transformation at one loop by comparing the 
amplitudes in the case of an electron and the  dyon that is its $SL(2,\mathbb{Z})$ image, and extend the invariance of squared amplitudes order by order in perturbation theory. We demonstrate that $S$-duality is property of all low-energy effective Abelian theories with electric and/or magnetic charges and see how the duality generically breaks down at high energies.
\end{quote}


\section{Introduction}

$S$-duality requires that the observables of a gauge theory are invariant under the transformation of the
the holomorphic gauge coupling,
\begin{equation}\label{tau}
\tau\equiv\frac{\theta}{2\pi}+\frac{4\pi i}{e^2},
\end{equation}
by $\tau \rightarrow -1/\tau$. Since $\theta$ is a periodic variable, the additional shift symmetry $\theta \rightarrow \theta + 2 \pi$ implies a full $SL(2,\mathbb{Z})$ duality. $S$-duality is directly established for ${\mathcal N}=4$ supersymmetric Yang-Mills\footnote{For ${\mathcal N}=4$ supersymmetric Yang-Mills $SL(2,\mathbb{Z})$ is  not just a duality, but an actual  invariance of the spectrum.}  theories \cite{Vafa} and free $U(1)$ gauge theories
\cite{Cardy1,Witten2,Lozano}. The situation for interacting $U(1)$'s is unclear since $S$-duality interchanges electric and magnetic charges which cannot be simultaneously included in a local, manifestly Lorentz invariant Lagrangian \cite{Dirac2,Hagen,Zwanziger:1971}. 

Here we will show that holomorphically normalized photon helicity amplitudes with $N_+$ positive helicity photons, 
$N_-$ negative helicity photons, and $I$ internal photon lines transform as modular forms under $SL(2,\mathbb{Z})$  of weight $(I+N_-,I+N_+)$, and that 
for canonically normalized photons the amplitude transforms by a phase independent of $I$, so that the magnitude of the amplitude is invariant. This means that perturbative amplitudes are mapped to perturbative amplitudes under duality. Moreover, the dual amplitudes can be verified by a perturbative calculation using the Zwanziger formalism \cite{Zwanziger:1971}, which introduces  a Lagrangian with local couplings for both electric and magnetic charges simultaneously.

In general $SL(2,\mathbb{Z})$ duality maps a particle with electric charge $q^\prime$ to a dual particle with electric charge $q$ and magnetic charge $g$.
Since, due to Dirac-Schwinger-Zwanziger charge quantization \cite{Dirac,Schwinger,Zwanziger:1968rs}, electric and magnetic charges have inverse coupling strengths (i.e. the magnetic fine structure constant is $\alpha_M\sim 1/\alpha$) it is often suggested that $S$-duality interchanges weak and strong coupling but we will see that this is not the case if one uses purely local couplings. Here we are using duality  in the sense of Seiberg duality \cite{Seiberg:1994pq} or the duality between AdS$_5$ and a 4D conformal field theory;  that is, duality is the occurrence of two different descriptions of the exactly the same physics. We will also show how $S$-duality is implemented in the Zwanziger formalism \cite{Zwanziger:1971} as a local field redefinition with a change of coupling constant, which means that $S$-duality is a property of any low-energy effective theory with electric and/or magnetic charges.

    Consider, for example, a theory with a light electron, of mass $m$, and a heavy magnetic monopole of mass $M$. If the electron is weakly coupled then the monopole will be strongly coupled. However we can estimate that the contributions to the low-energy photon scattering amplitude from the monopole will be suppressed by 
    \beq
    \frac{\alpha_M^2}{\alpha^2} \,\frac{m^2}{ M^2} \sim\left(\frac{m}{\alpha M}\right)^4~,
    \label{suppression}
\eeq
provided that $M$ is sufficiently large.  In the real world this requires that, if any monopole exists, it must be much heavier than 70 MeV for perturbation theory to be useful, which is certainly the case. $SL(2,\mathbb{Z})$ duality would fail even in this simple theory if the different contributions to the scattering amplitude picked up different phases under the duality transformation.  Perturbatively we will see that the phase only depends on the external photons, which must be the same for all contributions to the amplitude.  Whether this is the case non-perturbatively remains an open problem, but Eq.~(\ref{suppression}) provides an estimate of the error if the duality is broken by non-perturbative effects at the scale $M$ and duality is really only a low-energy approximate duality. Given a realistic bound on monopole masses of 1 TeV, the fractional error in photon scattering amplitudes would be less than about $2 \times 10^{-17}$. In other words, for energies far below the mass scale of strongly coupled monopoles we can use a reliable low-energy effective theory that is under perturbative control.  As we will see, the same is true when there are heavy, strongly coupled electrons and very light magnetic monopoles (or dyons): the low-energy effective theory of the gauge interactions of the monopoles/dyons is perturbative. 

This low-energy effective theory approach is certainly familiar from the seminal analysis by Seiberg and Witten \cite{SeibergWitten}, where they looked at the leading terms in the derivative expansion of ${\mathcal N}=2$ supersymmetric Yang-Mills with an $SU(2)$ gauge group.  In the low-energy theory there are only the $U(1)$ gauge multiplet, a BPS monopole, or a BPS dyon. The full theory has other electrically charged particles (eg. the massive gauge bosons and gauginos), but they are integrated out of the theory. Nevertheless, this low-energy effective theory proved to be extremely interesting, and the approximate  $SL(2,\mathbb{Z})$ duality played a key role in the analysis.

In the following sections we first briefly review $SL(2,\mathbb{Z})$ duality and the Zwanziger 
formalism\cite{Zwanziger:1971}. We then proceed to 
discuss photon helicity amplitudes and their duality transformations. We further explain how this analysis proceeds to 
all orders in perturbation theory in the effective theory. We then explore how $S$-duality can fail at high energies and finally apply our analysis to the Seiberg-Witten theory \cite{SeibergWitten}.


\section{$SL(2,\mathbb{Z})$ Duality Transformations}
Let us start with a
$U(1)$ gauge theory with coupling $e$ and non-vanishing $\theta$
angle, using a non-canonical (holomorphic) normalization of the
field strength:
\begin{equation}\label{free}
\mathcal{L}_{\text{free}}=-\frac{1}{4e^2}F_{\mu\nu}F^{\mu\nu}-\frac{\theta}{32\pi^2}F_{\mu\nu}\tilde{F}^{\mu\nu}~,
\end{equation}
where $F$ is the electromagnetic field strength and
$\tilde{F}_{\mu\nu}=\frac{1}{2}\epsilon_{\mu\nu\rho\sigma}F^{\rho\sigma}$
is the dual field strength.
Using the holomorphic gauge coupling,  $\tau$, 
the free-field Lagrangian may be written as
\beq
\label{freeholo}
\mathcal{L}_{\text{free}}=-\text{Im}\frac{\tau}{32\pi}(F_{\mu\nu}+i\tilde{F}_{\mu\nu})^2~.
\eeq
One can see that shifting $\tau$ by an integer $\tau\to\tau+n$ leads to a symmetry of
the theory, as this corresponds to a shift $\theta\to\theta+2\pi n$,
aka $T$-symmetry. Usually we perform the path integral over the gauge
 potential, $A_\mu$,  defined by
 \beq F_{\mu\nu}= \partial_\mu A_\nu- \partial_\nu A_\mu
 \eeq
 which can have a local coupling to electric currents.
 A change of variables in the path
integral can be performed so that we integrate over the dual potential \cite{Lozano} defined by
 \beq \tilde F_{\mu\nu}= \partial_\mu B_\nu- \partial_\nu B_\mu \eeq
which can have local couplings to magnetic currents, but is a non-local function of $A_\mu$.
 The form of the Lagrangian for $B_\mu$ is the same as (\ref{freeholo})
 with the replacement
$\tau\to-1/\tau$. This is not a symmetry of the theory but a duality, usually called
$S$-duality, which exchanges electric and magnetic fields with one
another. The $S$ and $T$ generators can be combined to obtain an
$SL(2,\mathbb{Z})$ group\footnote{Technically only the projective group
$PSL(2,\mathbb{Z})=SL(2,\mathbb{Z})/\{\pm 1\}$ acts on $\tau$, but we will ignore this distinction.} of dualities
\begin{equation}\label{sl2z}
\tau\to\tau'=\frac{a\tau+b}{c\tau+d},
\end{equation}
with $a,b,c,d$ integers satisfying $ad-bc=1$. Under such a
transformation, the electric current $J^\mu$ and the magnetic current $K^{\mu}$  transform as
\begin{equation}\label{currents}
\left(\begin{matrix} J^{\prime \mu } \\ K^{\prime \mu} \end{matrix}
\right) = \left(\begin{matrix}  a & - b\\-c & d \end{matrix}\right)\left(\begin{matrix} J^\mu \\ K^\mu \end{matrix} \right)~.
\end{equation}
This means that $SL(2,\mathbb{Z})$ duality maps a particle with
electric charge $q^\prime$ to a dual particle with electric charge $q$
and magnetic charge $g$, which is referred to as a dyon \cite{Schwinger:1969ib}.
While the matter fields are transformed by local field redefinitions the non-local transformation
of the gauge potential gives an air of mystery to $S$-duality, but we will see later that the 
mystery dissipates with the choice of a different Lagrangian formulation for the gauge field.

Given a theory with gauge coupling $e_d$ containing a dyon with electric charge $q$ and magnetic charge $g$ (in units of $e_d$ and $4 \pi/e_d$) we can perform 
an $SL(2,\mathbb{Z})$ transformation that takes us to a theory with
coupling $\tau^\prime$ and only an electric charge $q^\prime$, where
$q^\prime$ is the greatest common divisor of $q$ and $g$. This is done by choosing  $c=g/q^\prime$,
$d=q/q^\prime$, and where $a$, $b$ satisfy $a q- b g=q^\prime$. Then
we see from (\ref{currents}) that we have a mapping to the theory with
a electric charge $q^\prime$. From \eqref{sl2z} we see that the coupling $e$ satisfies
\beq
e^{2}=e_d^2\,|c \tau +d |^2~.
\label{dualcoupling}
\eeq

 For $U(1)$ theories with a CP-violating  vacuum angle
$\theta$,  Witten \cite{Witten} showed that the effective (low-energy) electric charge of a dyon is $Q=q+g\frac{\theta}{2\pi}$.
It is easily seen that   the Witten charge
$Q$ is $T$-invariant. In fact, the invariance of the Witten charge
restricts us to $SL(2,\mathbb{Z})$ rather than $SL(2,\mathbb{R})$.

By requiring the equations of motion to be covariant under
$SL(2,\mathbb{Z})$, we can extract the transformation properties of
the gauge field strength \cite{Csaki}. Maxwell's equations (incorporating the Witten charge) may be
written as
\begin{equation}\label{Maxwell}\frac{\text{Im}(\tau)}{4\pi}\partial_{\mu}(F^{\mu\nu}+i\tilde{F}^{\mu\nu})=J^{\nu}+\tau
K^{\nu}~.\end{equation}

The current \eqref{currents} and gauge coupling \eqref{sl2z}
transformations can be combined with the mappings
\begin{gather}\label{fieldtrans}
F'^{\mu\nu}+i\tilde{F}'^{\mu\nu}=\left(c\tau^*+d\right)\left(F^{\mu\nu}+i\tilde{F}^{\mu\nu}\right)~,\\
 F'^{\mu\nu}-i\tilde{F}'^{\mu\nu} =\left(c\tau+d\right)\left(F^{\mu\nu}-i\tilde{F}^{\mu\nu}\right) 
 \nonumber \end{gather}
to see that the Maxwell equations \eqref{Maxwell} are duality covariant. 

The form of this transformation makes it convenient to introduce helicity eigenstate field strengths
\cite{Silberstein}:
\begin{equation}\label{plusminus}
F_{\pm}^{\mu\nu}=F^{\mu\nu}\pm
i\tilde{F}^{\mu\nu}~.
\end{equation}
Since $F_+$ transforms as $(1,0)$ and $F_-$ as $(0,1)$ under the
Lorentz group, these represent the $\pm$ helicities of the
photon, which will allow us to easily make contact with spinor helicity techniques.
The mapping taking us from a dyon to an electric charge, described above  Eq.  \eqref{dualcoupling}, is equivalent to
\begin{equation}\label{helfieldtrans}
F_{\pm}'=D_{\mp}F_{\pm},
\end{equation}
where we define 
\begin{equation}
D_{\pm}\equiv\left(\begin{array}{c}c\tau+d\\ c\tau^* +d\end{array}\right)=\frac{\left(Q\pm ig/\alpha_d\right)}{q^\prime}~,
\end{equation}
and where $\alpha_d=e_d^2/4 \pi$.
From \eqref{fieldtrans} and \eqref{helfieldtrans} we see that $F_{+}^{\mu\nu}$ transforms under duality as a modular form \cite{Witten2} of weight (0,1) while 
$F_{-}^{\mu\nu}$ transforms under duality as a modular form of weight (1,0).

Since helicity eigenstate field strengths have simple modular transformations, it is
especially convenient (as can be seen in Appendix A) to use the spinor
helicity method of writing scattering amplitudes by decomposing Lorentz vectors and tensors into spinor
products \cite{Zee,Elvang}. 
It is straightforward to define the
polarization bispinors for gauge bosons: we want transversality
($\epsilon(k)\cdot k=0$) and they should be dimensionless. This forces
the polarization bispinor to be
\begin{equation}\label{polspinors}
\epsilon_{a\dot{a}}^-(k)=\sqrt{2}\frac{k_a p_{\dot{a}}}{[kp]},\quad\epsilon_{a\dot{a}}^+(k)=\sqrt{2}\frac{p_ak_{\dot{a}}}{\langle p k\rangle},
\end{equation}
where we are free to choose an arbitrary reference momentum $p$, which is a manifestation
of our freedom to choose a gauge. 
The duality transformations (\ref{helfieldtrans})  imply that the photon polarization (bispinor) transforms under duality by
\begin{equation}\label{polspinorstransf}
\epsilon_{a\dot{a}}^{\pm\prime}(k)= D_{\mp}\epsilon_{a\dot{a}}^\pm(k).
\end{equation}
This provides a simple method for directly obtaining dual photon helicity amplitudes.


\section{Zwanziger Formalism}
In order to check the $SL(2,\mathbb{Z})$ duality transformation of helicity amplitudes we will need to perturbatively calculate the  photon helicity amplitude generated by a dyon loop.
This calculation can performed using the Zwanziger two potential formulation \cite{Zwanziger:1971,Brandt:1978} of QED.  While there are still only 
two propagating photon degrees of freedom this formulation allows for local couplings to both electric and magnetic charges. This type of formulation was later rediscovered and generalized by Schwarz and Sen \cite{Schwarz:1993vs}.  By
introducing an extra gauge potential $B^{\mu}$, that couples to magnetic currants, and an arbitrary
four-vector $n^{\mu}$, Zwanziger showed that
the field strength and its dual may be written as
\begin{gather}\label{Zwanz}
F=\frac{1}{n^2}\left(n\wedge[n\cdot(\partial\wedge
  A)]-*\left\{n\wedge [n\cdot(\partial\wedge B)]\right\}\right),\\
\tilde{F}=\frac{1}{n^2}\left(*\left\{n\wedge[n\cdot(\partial\wedge
  A)]\right\}+n\wedge[n\cdot(\partial\wedge B)]\right),
\end{gather}
where $(a\wedge b)_{\mu\nu}=a_{\mu}b_{\nu}-a_{\nu}b_{\mu}$,
$(a\cdot G)^{\nu}=a_{\mu}G^{\mu\nu}=-G^{\nu\mu}a_{\mu}=-(G\cdot
a)^{\nu}$, and $a\cdot *(b\wedge
c)=a_{\mu}\epsilon^{\mu\nu}{}_{\rho\sigma}b^{\rho}c^{\sigma}$ for
four-vectors $a,b,c$ and antisymmetric tensor $G$. 
The generalization of Zwanziger's Lagrangian \cite{Zwanziger:1971} incorporating the $\theta$-angle \cite{Csaki} is:
\beq
{\mathcal L}&=&-{\rm Im} \frac{ \tau}{8\pi n^2} \left\{ \left[ n\cdot\partial \wedge (A+iB)\right]\cdot \left[ n\cdot \partial \wedge (A-iB)\right]\right\} \nonumber \\ &&-{\rm Re} \frac{ \tau}{8\pi n^2} \left\{ \left[ n\cdot\partial \wedge (A+iB)\right]\cdot \left[ n\cdot *\partial \wedge (A-iB)\right]\right\} \nonumber \\ &&-\Re[(A-iB)\cdot(J+\tau K)] .
\label{eq:fullZwanziger}
\eeq

\begin{figure}[h!]
\centering\includegraphics[width=\textwidth]{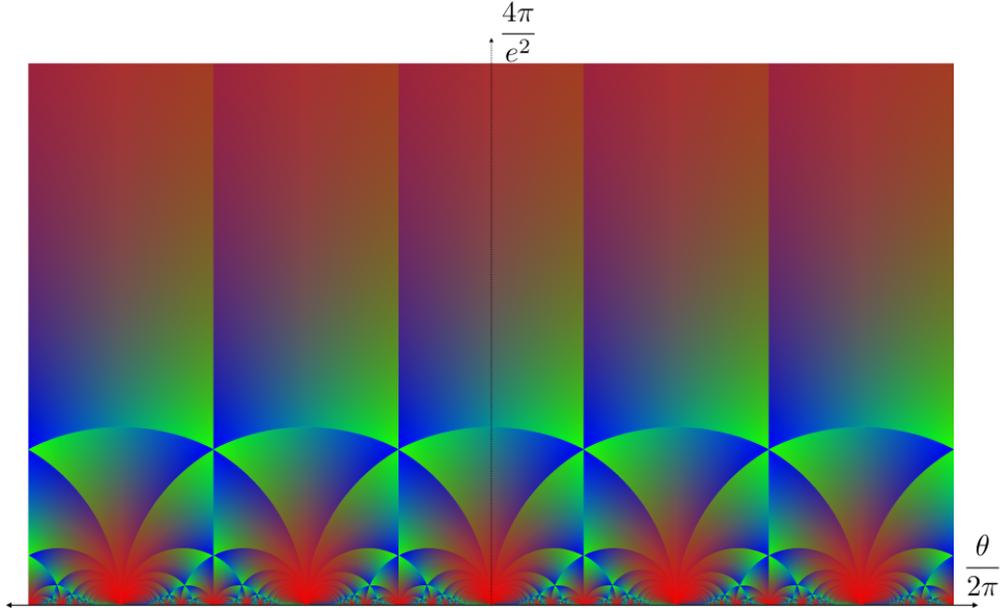}
\caption{Tiling of hyperbolic half-plane under the action of
  $SL(2,\mathbb{Z})$. The red, green, and blue shadings identify the
  mappings of each hyperbolic triangle's vertices, with the red vertex
at $\tau=i\infty$ corresponding to a weakly coupled electric theory. The other red regions correspond to the mapping
of this region to other dual, weakly coupled descriptions.}\label{HPpic}
\end{figure}

Using this Lagrangian with one species
of fermion and restoring
canonical normalization momentarily, we can anticipate the results of our loop calculations by examining the \emph{local} coupling
strength between a dyon of charge $(q,g)$ (in units of $e$ and $4
\pi/e$) and the complexified electromagnetic gauge potential, 
$A_\mu-iB_\mu$, which is
\begin{equation}
\includegraphics[width=.55\textwidth]{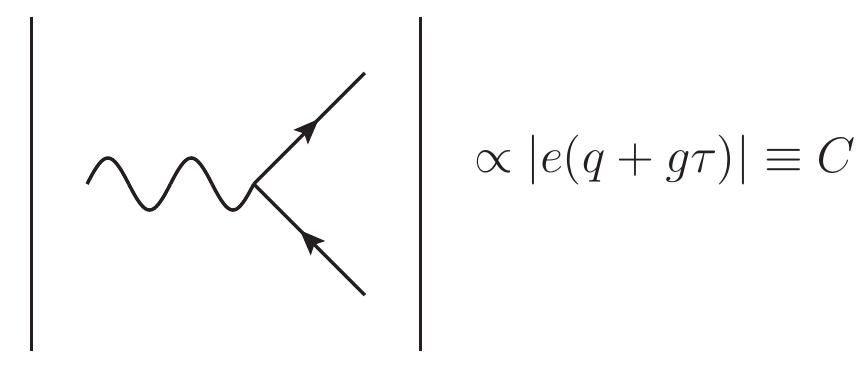}
\end{equation}
Because the Dirac-Schwinger-Zwanziger charge quantization condition forces monopoles to couple with the inverse of the electric coupling, one may be concerned that the magnetic fine structure constant
is too large to be an expansion parameter in a
perturbative calculation (which could be $S$-dual to a perturbative
electric theory). However, we can (making use of  \eqref{dualcoupling} and the requirement that $ad-bc=1$)  simply calculate the duality transformed local
coupling to the complexified dual gauge field:
\begin{align}\label{coupdual}
C' &= \left|e_d\,(q'+g'\tau')\right|\\
&=
e\left|c\tau+d\right|\left|(aq-bg)+(-cq+dg)\frac{a\tau+b}{c\tau+d}\right|\\
&= e\left|(ad-bc)(q+g\tau)\right|\\
&=e\left|q+g\tau\right|=C.
\end{align}
Thus the magnitude of the local coupling remains unchanged, in other words it is an $SL(2,\mathbb{Z})$ invariant. In particular, if we start with a
purely electric theory $(q',g')=(1,0)$ such that $C=e$ is small, then
the local coupling after a duality transformation will remain
small, and perturbative expansions are possible. For a given
perturbative theory there are an infinite set of mappings of the
holomorphic gauge coupling $\tau$ in the hyperbolic
half-plane\footnote{A half-plane since the gauge coupling is always
  real and positive.} $H$, and
these must all be weakly coupled theories.
The fundamental domain tiles $H$
with congruent hyperbolic triangles, as shown in Figure \ref{HPpic}, and we can see there is a complex pattern of weakly coupled theories.

\section{Photon-Photon Scattering}
At one-loop photons can scatter off other photons, the simplest process being
 $\gamma\gamma\to\gamma\gamma$. It has long been known \cite{Heisenberg,Karplus,Dicus} that the low-energy, one-loop effective field theory for QED includes a quartic photon interaction given by the
Euler-Heisenberg Lagrangian (using holomorphic normalization as in Eq. \eqref{free}, and dropping $^\prime$s  on field strengths for brevity)
\begin{equation}\label{EH}
\mathcal{L}_{\rm EH}=\frac{q^{\prime 4}}{360\,m^4\, 16\pi^2}[4(F^2)^2+7(F\tilde{F})^2],
\end{equation}
where $m$ and $q'$ are the mass and charge of the heavy fermion integrated out of the
theory. For energies much less than $m$,
the amplitudes for light-by-light scattering are most
easily calculated using the effective Euler-Heisenberg Lagrangian one helicity
configuration at a time. Labelling the amplitude by the
helicities of the (all incoming) photons, $\lambda_i=\pm1$, and going to canonically normalized fields, the helicity amplitudes are \cite{Karplus}:
\begin{gather}\label{helamps}
\mathcal{M}^{\pm\pm\mp\mp}=\frac{11\, \alpha^2\, q^{\prime 4}}{45\,m^4}s^2,\\
\mathcal{M}^{\pm\mp\pm\mp}=\frac{11\, \alpha^2\, q^{\prime 4}}{45\,m^4}t^2,\quad
\mathcal{M}^{\pm\mp\mp\pm}=\frac{11\, \alpha^2\, q^{\prime 4}}{45\,m^4}u^2,\\
\mathcal{M}^{\pm\pm\pm\mp}=\mathcal{M}^{\pm\pm\mp\pm}=\mathcal{M}^{\pm\mp\pm\pm}=\mathcal{M}^{\mp\pm\pm\pm}=0,\\
\mathcal{M}^{\pm\pm\pm\pm}=-\frac{\alpha^2\, q^{\prime 4}}{15\,m^4}(s^2+t^2+u^2),
\end{gather}
where $s,t,u$ are the usual Mandelstam invariants.
The factor of $e$ for each photon leg accounts for the canonical field normalization. 

Now consider the photon helicity amplitude generated by a dyon loop.
The calculation, using the Zwanziger formalism, is performed in Appendix A.
The easier method, which gives exactly the same answer, is to employ the duality transformations
\eqref{helfieldtrans}; note that
since $\tilde{F}^2=-F^2$, we may rewrite our Euler-Heisenberg
Lagrangian \eqref{EH} in the helicity eigenstate basis as
\begin{equation}\label{helEH}
\mathcal{L}_{EH}=\frac{q^{\prime 4}}{5760\,m^4\,16\pi^2}\left[22F_+^2F_-^2-3[(F_+^2)^2+(F_-^2)^2]\right].
\end{equation}
The duality transformations (\ref{helfieldtrans}) immediately give us
the dual Euler-Heisenberg  Lagrangian 
\begin{equation}\label{dyonL}
\begin{split}
\mathcal{L}_d&=\frac{1}{5760\,m^4\, 16\pi^2}\left\{22(Q^2+g^2/\alpha_d^2)^2F_+^2F_-^2\right.\\
&\quad\left.-3\left[(Q-ig/\alpha_d)^4(F_+^2)^2+(Q+ig/\alpha_d)^4(F_-^2)^2\right]\right\}.\end{split}
\end{equation}
That this is actually a proper 
Lagrangian can be seen by writing it in terms of the usual field strengths\footnote{This is the same result
(after taking $\theta\rightarrow 0$)
as equation (16) of ref. \cite{Kovalevich}, where a
classical Lorentz force law analogy is used to argue for this form.}:
\begin{equation}
\begin{split}
\mathcal{L}_d&=\frac{1}{360\,m^4\,16\pi^2}\left[4[(Q^2-g^2)^2+7Q^2g^2](F^2)^2\right.\\
&\quad\left.+[7(Q^2-g^2)^2+16Q^2g^2](F\tilde{F})^2-12Qg(Q^2-g^2)F^2(F\tilde{F})\right]~.
\end{split}
\end{equation}

Since the kinematics of photon scattering are unchanged, the effective dyon Lagrangian (\ref{dyonL}) yields the dual helicity amplitudes which match the  explicit dyon loop
calculation of Appendix A:
\begin{gather}
\mathcal{M}^{\pm\pm\mp\mp}_d=\frac{11\,\alpha_d^2\,(Q^2+g^2/\alpha_d^2)^2}{45\,m^4}s^2,\\
\mathcal{M}^{\pm\mp\pm\mp}_d=\frac{11\,\alpha_d^2\,(Q^2+g^2/\alpha_d^2)^2}{45\,m^4}t^2,\quad
\mathcal{M}^{\pm\mp\mp\pm}_d=\frac{11\,\alpha_d^2\,(Q^2+g^2/\alpha_d^2)^2}{45\,m^4}u^2,\label{helamps1}\\
\mathcal{M}^{\pm\pm\pm\mp}_d=\mathcal{M}^{\pm\pm\mp\pm}=\mathcal{M}^{\pm\mp\pm\pm}=\mathcal{M}^{\mp\pm\pm\pm}=0,\label{helamps2}\\
\mathcal{M}^{\pm\pm\pm\pm}_d=-\frac{\alpha_d^2\,(Q\mp ig/\alpha_d)^4}{15\,m^4}(s^2+t^2+u^2).\label{helamps3}
\end{gather}
We have included a factor of the dual gauge coupling $e_d$ for each photon leg to again return to canonical field normalization.
We can see that the parity, P, and time-reversal, T, preserving terms
depend only on the duality invariant $\alpha_d(Q^2+g^2/\alpha_d^2)$, and as expected
$\mathcal{M}^{++++}\neq\mathcal{M}^{----}$ due to 
P and T violation from the appearance of magnetic charge $g$ (a CP
pseudoscalar). Indeed, the appearance of imaginary terms in an 
amplitude would normally be contrary to the optical theorem in a
low-energy effective theory  devoid of dyon pair creation, but we
see that the optical theorem is indeed satisfied:
\begin{equation}
\mathcal{M}_d^{++++}-(\mathcal{M}_d^{----})^*=0.
\end{equation}

We can also see that for a fixed set of helicities that the dual amplitudes (\ref{helamps2}) are not equal to the original amplitudes (\ref{helamps}). However if the duality 
is to hold it must be the case that observables (i.e. the squares of amplitudes) are duality invariant. This is
indeed the case since
\begin{equation}
e_d^8 \left| (Q\pm i g/\alpha_d)^4 \right|^2= e_d^8 \left| (Q^2 + g^2/\alpha_d^2)^2 \right|^2= e^8 q^{\prime 8}
\end{equation}
where we have used Eq. (\ref{dualcoupling}) for the last equality. Thus, perhaps remarkably to some, duality invariance is true at leading order  in the loop expansion.


Making use of the helicity formalism lets us go even further. The
choices for polarization bi-spinors \eqref{polspinors} allow
helicity-specific decompositions 
\begin{equation}
F^{\pm}_{a\dot{a}b\dot{b}}=\sigma_{a\dot{a}}^{\mu}\sigma_{b\dot{b}}^{\nu}F_{\mu\nu}^{\pm}\,\Rightarrow\,
F^+_{a\dot{a}b\dot{b}}=-2\sqrt{2}k_{\dot{a}}k_{\dot{b}}\epsilon_{ab},\quad
F^-_{a\dot{a}b\dot{b}}=-2\sqrt{2}k_ak_b\epsilon_{\dot{a}\dot{b}}.
\end{equation}
We can invert these to find Lorentz products of field strengths, 
like those that appear in the effective Lagrangian (and for the polarizations in the corresponding amplitudes):
\begin{equation}\label{helcontract}
F_{\mu\nu}^+F^{+\mu\nu}=4[k_ik_j]^2,\quad
F_{\mu\nu}^-F^{-\mu\nu}=4\langle k_ik_j\rangle^2,\quad
F_{\mu\nu}^+F^{-\mu\nu}=0.
\end{equation}
The relation $\tilde{F}^{\pm}=\tilde{F}\pm i F=\mp iF^{\pm}$
gives the other contractions.
With the simple form of these Lorentz products, we can write an
expression for the full low-energy helicity amplitude of $N$ photons
from a fermion at one-loop. The integral representation of the
Euler-Heisenberg Lagrangian at one-loop is \cite{Dicus}
\begin{equation}\label{fullEH}
\mathcal{L}_{EH}=-\frac{1}{8\pi^2}\int_0^{\infty}\frac{dT}{T}e^{-m^2T}\left[\frac{q^{\prime 2}ab}{\tanh(q^\prime aT)\tan(q^\prime bT)}-\frac{q^{\prime 2}}{3}(a^2-b^2)-\frac{1}{T^2}\right],
\end{equation}
where $T$ is the proper time of a fermion with charge $q^\prime$, mass $m$, and 
\begin{equation}
a^2=\frac{1}{4}\sqrt{(F^2)^2+(F\tilde{F})^2}+\frac{1}{4}F^2,\quad
b^2=\frac{1}{4}\sqrt{(F^2)^2+(F\tilde{F})^2}-\frac{1}{4}F^2.
\end{equation}
Martin et. al. \cite{Martin} showed that this gives the full amplitude for $N$
photons ($N_+$ of which have positive helicity and $N_-=N-N_+$ with negative
helicity) which can be written as
\begin{equation}
\mathcal{M}^{N_+;N_-}=-\frac{m^4}{8\pi^2}\left(\frac{2\,q^\prime\,e}{m^2}\right)^{N_+ +N_-} c_{sp}(N_+,N_-)\chi^+\chi^-,
\end{equation}
with coefficients $c_{sp}$ and spinor products $\chi^{\pm}$ defined by
\begin{gather}
c_{sp}(N_+,N_-)=(-1)^{N/2}(N-3)!\sum_{k=0}^{N_+}\sum_{j=0}^{N_-}(-1)^{N_- -j}\frac{B_{k+j}B_{N-k-j}}{k!j!(N_+-k)!(N_--j)!}\\
\chi^+=\frac{(N_+/2)!}{2^{N_+/2}}\left([12]^2[34]^2\cdots[(N_+-1)N_+]^2+\text{all
  permutations}\right)\\
\begin{split}\chi^-&=\frac{(N_-/2)!}{2^{N_-/2}}\left(\langle(N_+ +1)(N_+ +2)\rangle^2\langle(N_+ +3)(N_+ +4)\rangle^2\right.\\
&\quad\left.\cdots\langle(N-1)N\rangle^2+\text{all perms.}\right)\end{split}
\end{gather}
where $B_{2n}$ are the Bernoulli numbers. This expression
is valid at leading order in the derivative expansion.

To dualize the theory to one of dyons, we note that \eqref{polspinorstransf} implies that
$\chi^{\pm}\to D_{\mp}^{N_\pm}\chi^{\pm}$, so that a dual
amplitude is
\begin{equation}
\mathcal{M}_d^{N_+;N_-}=-\frac{m^4}{8\pi^2}\left(\frac{2\,e_d}{m^2}\right)^N c_{sp}(N_+,N_-)D_-^{N_+}\,D_+^{N_-}\chi^+\chi^-.
\label{generalamp}
\end{equation}
We can see that the holomorphically normalized amplitude transforms under $SL(2,\mathbb{Z})$ duality as a modular form of weight $(N_-,N_+)$.


\section{Higher Orders}

To go beyond one-loop requires additional information, since
the two-loop diagram has one internal photon line. Since the numerator of the photon propagator contains
\begin{equation}
\sum_\lambda \epsilon_{a\dot{a}}^{*\lambda}(k) \epsilon_{a\dot{a}}^\lambda(k)~,
\end{equation}
one might think that all we need is an additional rescaling by 
\begin{equation}
D^*_{\pm} D^{\vphantom{*}}_{\pm} =\left(\begin{array}{c}|c\tau+d|^2\\ |c\tau^* +d|^2 \end{array}\right)~.
\end{equation}
This can be verified by examining the source-gauge coupling term of
\eqref{eq:fullZwanziger}: since $J^{\mu}+\tau K^{\mu}$ has a known
transformation under duality, so too must $A_{\mu}\pm iB_{\mu}$ for
the Lagrangian to remain duality covariant. It is just this transformation,
\begin{gather}\label{gaugetrans}
A^\prime_{\mu}+iB^\prime_{\mu}=\left(c\tau^*+d\right)\left(A_{\mu}+ iB_{\mu}\right)~,\\
A^\prime_{\mu}-iB^\prime_{\mu} =\left(c\tau+d\right)\left(A_{\mu}- iB_{\mu}\right)~,
 \nonumber \end{gather}
  that agrees \cite{Csaki} with the transformation of the field strength \eqref{fieldtrans}. 
  This means that  $S$-duality can be implemented \cite{Schwarz:1993vs} as a local field redefinition of the gauge and matter fields along with a transformation of the coupling.
  
  In order to see the range of applicability of $S$-duality, consider a generic low-energy effective $U(1)$ gauge theory  where one species of  electrically and/or magnetically charged particles is light enough to be included in the low-energy dynamics as point-like particles. Specifically this means that the Compton wavelength ($\lambda \sim$ 1/mass) is much longer that the physical size. In the case of a  `t Hooft-Polyakov monopole \cite{'tHooft:1974qc} the size is $\sim 1/(e\, v)$, where $v$ is the VEV the breaks the non-Abelian gauge symmetry down to $U(1)$.  In this case, $\lambda \gg 1/(e\, v)$, we can treat the monopoles as point charges. Since $U(1)$ theories are infrared free, if the mass is sufficiently small compared to $v$, then the coupling will be perturbative at low-energies, and we can use the Zwanziger Lagrangian as the low-energy effective theory. As we have seen, this effective theory must enjoy $S$-duality to a good approximation.
  
  Returning to our perturbative argument, at the level of the gauge
propagator, we see that each internal photon line will
contribute a factor $D^*_{\lambda}D_{\lambda}=|c\tau+d |^2$ to the dual amplitude. Thus a holomorphically normalized amplitude with $I$ internal lines transforms under $SL(2,\mathbb{Z})$ duality as a modular form of weight $(I+N_-,I+N_+)$. For canonically normalized amplitudes 
we see, using Eq. \eqref{dualcoupling}, that the factor of $D^*_{\lambda}D_{\lambda}$ simply converts gauge couplings on the internal line to the dual couplings (the phases cancel between $D^*_{\lambda}$ and $D_{\lambda}$), so the higher order amplitude transforms by the same phase as the leading order amplitude, and the squares of the amplitudes are invariant order by order in perturbation theory. 

So far we have only discussed photon scattering, but low-energy scattering involving fermions or scalars follows a similar story. Tree level scattering enjoys $S$-duality because it
is a property of the classical theory. Adding external photon lines generates a $SL(2,\mathbb{Z})$ relative phase between the dual amplitudes as described above, while internal photons add no additional phase factor.  Thus, in a given duality basis, there is no relative phase between the leading term and the higher order terms, so again squares of canonically normalized amplitudes are duality invariant. However as we will see in the next section, $S$-duality can break down when sufficiently hard photons are involved.


\section{High Energy Breakdown}

So far we have only worked at leading order in the derivative
expansion where the amplitudes with only one $+$ helicity (or only one
$-$ helicity) vanished. This is not true at next-to-leading order in
the 1-loop derivative expansion \cite{Gusynin}:
\begin{gather*}
\mathcal{M}_{(1)}'^{\pm\pm\mp\mp}=-\frac{4\,\alpha^2\,q'^4}{315\,m^6}s^3,\quad
\mathcal{M}_{(1)}'^{\pm\mp\pm\mp}=-\frac{4\,\alpha^2\,q'^4}{315\,m^6}t^3,\quad
\mathcal{M}_{(1)}'^{\pm\mp\mp\pm}=-\frac{4\,\alpha^2\,q'^4}{315\,m^6}u^3\\
\mathcal{M}_{(1)}'^{\pm\pm\pm\mp}=\mathcal{M}_{(1)}'^{\pm\pm\mp\pm}=\mathcal{M}_{(1)}'^{\pm\mp\pm\pm}=\mathcal{M}_{(1)}'^{\mp\pm\pm\pm}=-\frac{\alpha^2\,q'^4}{315\,m^6}stu\\
\mathcal{M}_{(1)}'^{\pm\pm\pm\pm}=\frac{2\,\alpha^2\,q'^4}{63\,m^6}stu
\end{gather*}
At higher orders in the derivative expansion, the duality relations of amplitudes goes through as above. At next-to-leading order, the dual amplitude for the new helicity configuration is 
\beq
\mathcal{M}_d'^{\pm\pm\pm\mp}=-\frac{\alpha_d^2 (Q\mp i g/\alpha_d)^3 (Q\pm i g/\alpha_d)stu}{315\,m^6}
\eeq
The amplitude-squared is again duality invariant.

While it would seem that $S$-duality could continue to hold to higher and higher order in the derivative expansion, this is not the case, and the reason is somewhat subtle.
The Witten charge of the dyon, $Q=q+g\theta/2\pi$, is only correct when the charge is probed by a low-energy photon.
The extra $\theta$ dependent part of the charge is spread out in fermion zero-modes over a region of size $\sim m^{-1}$, where
$m$ is the mass of the lightest electrically charged fermion \cite{Callan:1982au}. Sufficiently high energy
photons, $E \gg m$, can resolve the core and the zero-mode cloud separately. A high energy photon that resolves the core of the dyon will simply couple to the charge $q$. 

This means that at high energies the Zwanziger effective Lagrangian breaks down, since the electric photon coupling has
a form factor with a scale dependence set by the mass of the lightest electrically charged fermion. At the very least we would need to include higher dimension operators, suppressed by the the scale $m$, that account for the low-momentum behavior of the form factor. In general we would only expect $S$-duality to hold exactly in a theory with an 
$SL(2,\mathbb{Z})$ invariant spectrum.

For a low-energy theory of electrons  (valid far below the mass, $M$, of the lightest monopole)  there are no form factors present in the effective theory, so we can use our analysis of higher loop corrections as in the previous section.  In the dual description where the weakly coupled electron is mapped to a weakly coupled monopole, the effects of form factors only appear for photons with energies above $M$, so they are again irrelevant in the low-energy effective theory, and we can again proceed with our analysis of higher loop corrections as before.

When we include high energy photons however, the dual couplings will not, in general, satisfy (\ref{coupdual}) and the amplitudes calculated from the dual Lagrangian will no longer provide the correct phases (\ref{generalamp}) to ensure $SL(2,\mathbb{Z})$ duality.

\section{Seiberg-Witten Theory}

The most fully understood example of low-energy $S$-duality is the Seiberg-Witten theory \cite{SeibergWitten}. The theory is an ${\cal N}=2$ SUSY 
theory with an $SU(2)$ gauge group.  The VEV of the adjoint scalar, $\phi$, breaks $SU(2)$ down to $U(1)$, so  `t Hooft-Polyakov monopoles \cite{'tHooft:1974qc} appear in this theory, and they are, in fact, BPS states.
At particular points in the moduli space a monopole or a dyon becomes massless.
Parameterizing the moduli space by the gauge invariant $u={\rm Tr } \phi^2$, the masses of the two BPS states are given by
\beq
m(q,g;u)= \sqrt{2} \left| q \,a(u)+g \,a_D(u)\right|
\eeq
where, at leading order in the derivative expansion, $a$ and $a_D$ are given by hypergeometric functions
\beq
a(u)&=&- \sqrt{2(\Lambda^2+u)} \, F\left(-\frac{1}{2},\frac{1}{2},1;
\frac{2}{1+\frac{u}{\Lambda^2}} \right)~,\\
a_D(u)&=&
-i \frac{1}{2}\left(\frac{u}{\Lambda}-\Lambda\right)F\left(\frac{1}{2},\frac{1}{2},2;\frac{1} {2} \left(1-\frac{u}{\Lambda^2}\right)  \right).
\eeq
The monopole mass vanishes at $u=\Lambda^2$; Taylor expanding about this point\footnote{Taking $u$ to be real and $u>\Lambda^2$ for simplicity.} we have
\beq
m\approx \frac{u-\Lambda^2}{\sqrt{2}\Lambda}~.
\eeq
The holomorphic coupling is given by
 \beq
 \tau=\frac{\partial a_D/\partial u}{\partial a/\partial u}~.
 \label{Chap13:taupartial2}
  \eeq
So for a light monopole ($m \ll \Lambda$) we find, expressing $u$ in terms of $m$ and $\Lambda$,
\beq
\tau = i\frac{\pi}{\log \Lambda/m}~.
\label{alpha_d}
\eeq 
In the $S$-dual frame, where the monopole is mapped to an electric charge, $\alpha_d=i/\tau_d=-i\tau$. As we move on the moduli space, the mass $m$ changes. We see from (\ref{alpha_d}) that the dual electric coupling, $\alpha_d$, approaches zero as we approach $m=0$, which is
simply the ordinary perturbative running of the infrared free $U(1)$ coupling.
In the original frame, where the light state has a magnetic charge, it would seem that the coupling is very strong ($\alpha=1/\alpha_d$), however this is a statement about the electric coupling, and there are no light electrically charged particles in this low-energy effective theory.  In the usual formulation the monopole has only a non-local coupling \cite{Dirac2}, and it is not clear what we even mean by a coupling.  In the local Zwanziger formulation we see that the monopole has a small coupling, since it couples with strength $1/e=e_d$. The  unusual running of the gauge coupling ($\log$ rather than $1/\log$) has been explicitly verified in the Zwanziger formalism \cite{Laperashvili:1999pu}, but it comes as no surprise
since the inverse relation between the two couplings must be independent of renormalization scale, since it is required by Dirac-Schwinger-Zwanziger charge quantization \cite{Coleman:1982cx}. The fact that the $\beta$ function changes sign can be seen directly from the $SL(2,\mathbb{Z})$  transformation of the vacuum polarization \cite{Csaki,Argyres:1995jj}, in the case of transforming a monopole to an electron the phase is just $-1$. To see this explicitly we note that there are two (holomorphic) vacuum polarization amplitudes, $\mathcal{M}^{++}$ and $\mathcal{M}^{--}$ which transform as
\beq
\mathcal{M}^{++}= (c \tau^* +d)^2  \, \mathcal{M}_{d}^{++}~, \quad \mathcal{M}^{--}= (c \tau +d)^2  \, \mathcal{M}_{d}^{--}~.
\eeq
The transformation that takes a monopole $(q=0, g)$ to a electron $(q^\prime=g,0)$ has $c=g/q^\prime=1$ and $d=0$, so (setting $\theta=0$ for simplicity\footnote{See \cite{Csaki,Argyres:1995jj} for details of the full story with non-zero $\theta$.}) we have
\beq
\mathcal{M}^{++}= - \, \mathcal{M}_{d}^{++}~, \quad \mathcal{M}^{--}= - \, \mathcal{M}_{d}^{--}~,
\eeq
and we see that the $\beta$ function flips sign.
This discussion again makes it clear that $S$-duality, unlike Seiberg
duality \cite{Seiberg:1994pq}, interchanges weakly coupled, local theories with other weakly coupled, local theories.

\section{Conclusion}
We have shown that $S$-duality implies that holomorphically normalized photon helicity amplitudes should transform  as modular forms.  This means that canonically normalized amplitudes transform by just a phase.
  Using the Zwanziger formalism we were able to verify this at one-loop and to show how the structure of the photon propagator ensures that $S$-duality is preserved at higher loop order in the low-energy effective theory. This required showing that all contributions to a particular amplitude transform by an identical phase. The fact that $S$-duality can be implemented by a local field redefinition and a change of coupling constant in the 
Zwanziger formalism shows that $S$-duality works for any low-energy effective theory of electric and/or magnetic charges. We also saw how $S$-duality can fail at high energies once we are able to probe inside the zero-mode cloud that surrounds a magnetic charge. 
It would be very interesting to study the Seiberg-Witten theory at higher orders in the derivative expansion to see how the approximate $S$-duality of the effective theory begins to break down at higher energies since ${\mathcal N}=2$ theories do not have $SL(2,\mathbb{Z})$ invariant spectra. This should be especially important  near Argyres-Douglas points \cite{Argyres:1995jj} where both monopoles and dyons become light at the same point in the moduli space.

\appendix
\section*{Acknowledgments}

We thank C. Csaki, B. Heidenreich, N. Weiner, and E. Witten for useful discussions.
J.T. thanks the Aspen Center for Physics where part of this work was completed.
This work was supported in part by DOE under grant DE-SC-000999.


\numberwithin{equation}{section}

\setcounter{equation}{0}
\setcounter{footnote}{0}


\section{Dyon Loop Calculation}

It is most convenient to use the spinor helicity method to calculate photon helicity amplitudes \cite{Zee,Elvang}. This is done with the local isomorphism
$SO(1,3)=SU(2)_L\times SU(2)_R$; a null
four-vector $p^{\mu}$ may be written as the outer product of two commuting
spinors:
\begin{equation}
p_{a\dot{a}}=p_{\mu}\sigma^{\mu}_{a\dot{a}}=\sgn(p^0) p_ap_{\dot{a}},
\end{equation}
where (un)dotted indices label which $SU(2)$ subalgebra the
spinor $p_a,p_{\dot{a}}$ belong to. Since $p^2=\det
p_{a\dot{a}}=0$, we have that $p_{a\dot{a}}$ is rank one, which may
always be written as an outer product. We can form invariants by contracting spinor indices using  $\epsilon_{ab}$ and
$\epsilon_{\dot{a}\dot{b}}$ for $SU(2)_L$ and $SU(2)_R$. For two null vectors $p_{a\dot{a}}=p_ap_{\dot{a}}$,
$q_{a\dot{a}}=q_aq_{\dot{a}}$, we introduce the standard spinor products
\begin{equation}\label{spinprods}
2p\cdot
q=(\epsilon^{ab}p_aq_b)(\epsilon^{\dot{a}\dot{b}}p_{\dot{a}}q_{\dot{b}})\equiv\langle
qp\rangle[pq].
\end{equation}
Here we can see that these forms are actually Lorentz invariant: for
real momenta, $p_{\dot{a}}=(p_a)^*$, so that $\langle pq\rangle=[pq]^*$, implying that each is the square root of $2p\cdot q$,
up to a phase. 


As described above, the polarization bispinor for a photon of momentum $k$, $\epsilon_{a\dot{a}}^-(k)$ in \eqref{polspinors},
is defined with respect to an arbitrary momentum $q$.
The gauge transformation $q_a\to q_a+\eta k_a$ ($\eta$ an
arbitrary constant) shifts the polarization 
$\epsilon_{a\dot{a}}\to\epsilon_{a\dot{a}}+\eta
k_ak_{\dot{a}}$ ($\epsilon^{\mu}\to\epsilon^{\mu}+\eta k^{\mu}$), just like we
expect for a usual gauge transformation. As we are free to choose the
reference momentum $q$ independently for each polarization vector, it
is convenient to take one of the positive helicity momenta as
reference for the negative helicity polarizations, and vice
versa\footnote{This procedure has made calculating tree-level QCD helicity
amplitudes more of a joy than a pain \cite{Dixon}.}.

While the
matter sector in \eqref{eq:fullZwanziger} is identical to that of 
standard QED for fermionic dyons, the photon propagator and coupling
vertex become 
\begin{gather}
D^{ab}_{\mu\nu}(k)=\frac{-i}{k^2}\left[\left(g_{\mu\nu}-\frac{k_{\mu}n_{\nu}+n_{\mu}k_{\nu}}{n\cdot
    k}\right)\delta^{ab}+\left(\frac{\epsilon_{\mu\nu\rho\sigma}n^{\rho}k^{\sigma}}{n\cdot
    k}\right)\epsilon^{ab}\right],\\
(\Gamma_i^a)^{\mu}=iq_i^a\gamma^{\mu},
\end{gather}
where we observe both a standard diagonal term ($\delta^{ab}$) in the
charge space, $q_i^a=(Q_i,g_i/\alpha_d)$, as well as an off-diagonal term ($\epsilon^{ab}$),
which leads to dynamical mixing of the two gauge potentials.  Here
we assume that there is only one species of dyon; in general one would
sum over $i$. 

For plane-wave gauge fields, we require \cite{Deans,Brandt:1978} the
electric and magnetic external polarizations to obey
$\epsilon^{\mu}_a=(\epsilon^{\mu},\bar{\epsilon}^{\mu})$, with
\begin{equation}\label{magpols}
\bar{\epsilon}^{\mu}=-\frac{\epsilon^{\mu\nu\rho\sigma}\epsilon_{\nu}n_{\rho}k_{\sigma}}{n\cdot k}.
\end{equation}
This magnetic photon polarization vector is realized by either
appending an $A/B$ transition for one of the external photons, or by
enforcing the the free space duality constraint
\begin{equation}
n\cdot(\partial\wedge A)=-n\cdot *(\partial\wedge B)
\end{equation}
in an axial gauge. 

The appearance of the Lorentz symmetry breaking 
vector $n^{\mu}$ in the propagator and magnetic polarization vectors is
troubling. However, 
we have gauge symmetry at our disposal, and in fact different choices
of $n^{\mu}$ amount to a different choice of gauge. This is well-known for
space-like $n^{\mu}$, where the gauge transformation corresponds to
the solid angle subtended by the rotated Dirac string
\cite{Jackson}. Fixing only
spatial components of the gauge fields leads to 
straightforward canonical quantization, while using a lightcone gauge
makes the spinor decomposition of polarization vectors manifest (but
requires positive and negative helicity photons to be in different gauges). We
can have both qualities by employing the space-cone gauge \cite{Siegel}:
complexify $n$ so that
\[n^2=0,\quad n\cdot n^*>0,\]
where we have both the spacelike and null features we desire. The
holomorphic nature of the Lagrangian is preserved by ensuring that
$\mathcal{L}$ is independent of $n^*$. All fields are now in one gauge
defined by 
\[n_{a\dot{a}}=n_a^+n^-_{\dot{a}}=|+\rangle[-|,\]
orthogonal to two external momenta if we choose our reference spinors
$n_a^+=(k_i^+)_a$, 
$n_{\dot{a}}^-=(k_j^-)_{\dot{a}}$. This is essentially a Wick rotation
of the lightcone, equivalent to employing two real null vectors
\[n_{a\dot{a}}^{\pm}=n_a^{\pm}n_{\dot{a}}^{\pm}=|\pm\rangle[\pm|\]
as two lightcone gauges.

The spacecone gauge is a convenient simplification, but the true
independence of $F_{\mu\nu}$ from $n^{\mu}$ can be 
seen by examining the source of the Zwanziger identity
\eqref{Zwanz}. If we introduce the operators \cite{Ripka}
\begin{gather}
G_{\mu\nu\alpha\beta}=g_{\mu\alpha}g_{\nu\beta}-g_{\mu\beta}g_{\nu\alpha},\\
 K_{\mu\nu\alpha\beta}=\frac{1}{n^2}(n_{\mu}n_{\alpha}g_{\nu\beta}-n_{\mu}n_{\beta}g_{\nu\alpha}+n_{\nu}n_{\beta}g_{\mu\alpha}-n_{\nu}n_{\alpha}g_{\mu\beta}),\\
E_{\mu\nu\alpha\beta}=\frac{1}{4}\epsilon_{\mu\nu\rho\sigma}K^{\rho\sigma\gamma\delta}\epsilon_{\gamma\delta\alpha\beta}=\epsilon_{\mu\nu\rho\sigma}\frac{1}{n^2}g^{\sigma\gamma}n^{\rho}n^{\delta}\epsilon_{\gamma\delta\alpha\beta}
\end{gather}
with the convention that for contractions of these tensor operators we use
\begin{equation}
\mathcal{O}F\equiv\frac{1}{2}\mathcal{O}_{\mu\nu\rho\sigma}F^{\rho\sigma},\quad
\mathcal{O}_1\mathcal{O}_2=\frac{1}{2}\mathcal{O}_{1\mu\nu\alpha\beta}\mathcal{O}_2^{\alpha\beta\rho\sigma},
\end{equation}
then \eqref{Zwanz} is just the statement that $F=GF=(K-E)F$.
Since $G$ is independent of $n^{\mu}$, we know that the combination
$K-E$ must be as well, and ultimately varying $n^{\mu}$ will amount to
a choice of axial gauge. Thus even null $n^{\mu}$ may be employed in
\eqref{Zwanz} as a limiting case. 

The Zwanziger algebra itself is generated by
the elements $G$, $\epsilon$, and $K$, with the relations $K-\epsilon
K\epsilon=G$, $K^2=K$, and $\epsilon^2=-G$. While the structure of
this algebra may be worth pursuing in future work, we will not pursue it here.

Now we may rewrite quantities with Lorentz indices in terms of pairs of
spinor indices; of particular use are the Levi-Civita contractions
\begin{gather}
\epsilon_{a\dot{a}b\dot{b}c\dot{c}d\dot{d}}=\epsilon^{\mu\nu\rho\kappa}\sigma_{\mu
  a\dot{a}}\sigma_{\nu b\dot{b}}\sigma_{\rho c\dot{c}}\sigma_{\kappa
  d\dot{d}}
=4i(\epsilon_{ad}\epsilon_{bc}\epsilon_{\dot{a}\dot{b}}\epsilon_{\dot{c}\dot{d}}-\epsilon_{ab}\epsilon_{cd}\epsilon_{\dot{a}\dot{d}}\epsilon_{\dot{b}\dot{c}})\\
\epsilon^{\dot{a}a\dot{b}b\dot{c}c\dot{d}d}=\epsilon^{\mu\nu\rho\kappa}\bar{\sigma}_{\mu}^{
  \dot{a}a}\bar{\sigma}_{\nu}^{\dot{b}b}\bar{\sigma}_{\rho}^{\dot{c}c}\bar{\sigma}_{\kappa}^{\dot{d}d}=4i(\epsilon^{ad}\epsilon^{bc}\epsilon^{\dot{a}\dot{b}}\epsilon^{\dot{c}\dot{d}}-\epsilon^{ab}\epsilon^{cd}\epsilon^{\dot{a}\dot{d}}\epsilon^{\dot{b}\dot{c}})\\
\epsilon_{\mu\nu\rho\kappa}\sigma^{\mu}_{a\dot{a}}\sigma^{\nu}_{b\dot{b}}\bar{\sigma}^{\rho
  \dot{c}c}\bar{\sigma}^{\kappa
  \dot{d}d}=4i(\epsilon_{ab}\epsilon^{cd}\delta_{\dot{a}}^{\dot{c}}\delta_{\dot{b}}^{\dot{d}}-\epsilon_{\dot{a}\dot{b}}\epsilon^{\dot{c}\dot{d}}\delta_a^c\delta_b^d)
\end{gather}
and so on. This allows decomposition of magnetic polarization vectors
\eqref{magpols} and Lorentz products into spinor products:
\begin{gather}\label{lorentzprods}
\bar{\epsilon}_{a\dot{a}}^-=i\sqrt{2}\left(\frac{k_aq_{\dot{a}}}{[kq]}+\frac{k_ak_{\dot{a}}[qn]}{[kq][nk]}\right),\quad
\bar{\epsilon}_{a\dot{a}}^+=-i\sqrt{2}\left(\frac{q_ak_{\dot{a}}}{\langle
  qk\rangle}+\frac{k_ak_{\dot{a}}\langle nq\rangle}{\langle
  qk\rangle\langle kn\rangle}\right)\\
\epsilon_i^+\cdot\epsilon_j^+=\frac{\langle
  q_iq_j\rangle[k_ik_j]}{\langle q_ik_i\rangle\langle k_jq_j\rangle},\quad
\epsilon_i^-\cdot\epsilon_j^-=\frac{\langle k_ik_j\rangle[q_iq_j]}{[q_ik_i][k_jq_j]},\quad
\epsilon_i^-\cdot\epsilon_j^+=\frac{\langle
  k_iq_j\rangle[q_ik_j]}{[k_iq_i]\langle q_jk_j\rangle}\\
\epsilon_i^+\cdot
k_j=\frac{\langle q_ik_j\rangle[k_ik_j]}{\sqrt{2}\langle k_iq_i\rangle},\quad
\epsilon_i^-\cdot k_j=\frac{\langle
  k_ik_j\rangle[q_ik_j]}{\sqrt{2}[k_iq_i]},\quad
\epsilon_i^+\cdot\epsilon_i^-=1.
\end{gather}
Under the definitions
\begin{gather}\label{magpolhel}
z=\frac{\langle nq\rangle}{\langle
  qk\rangle\langle
  kn\rangle},\quad\bar{z}=\frac{[nq]}{[qk][kn]}\\
\bar{\epsilon}^+=-i(\epsilon^++\sqrt{2}zk),\quad
\bar{\epsilon}^-=i(\epsilon^--\sqrt{2}\bar{z}k),
\end{gather}
we see that the magnetic polarization bispinors of specific helicity
are essentially $\pm i$ times the corresponding electric bispinors, up
to a gauge transformation shift $z$ or $\bar{z}$.


At leading order, light-by-light scattering can be computed using the
Zwanziger Feynman rules. The three diagrams that contribute are: 
\begin{center}\includegraphics[width=\textwidth]{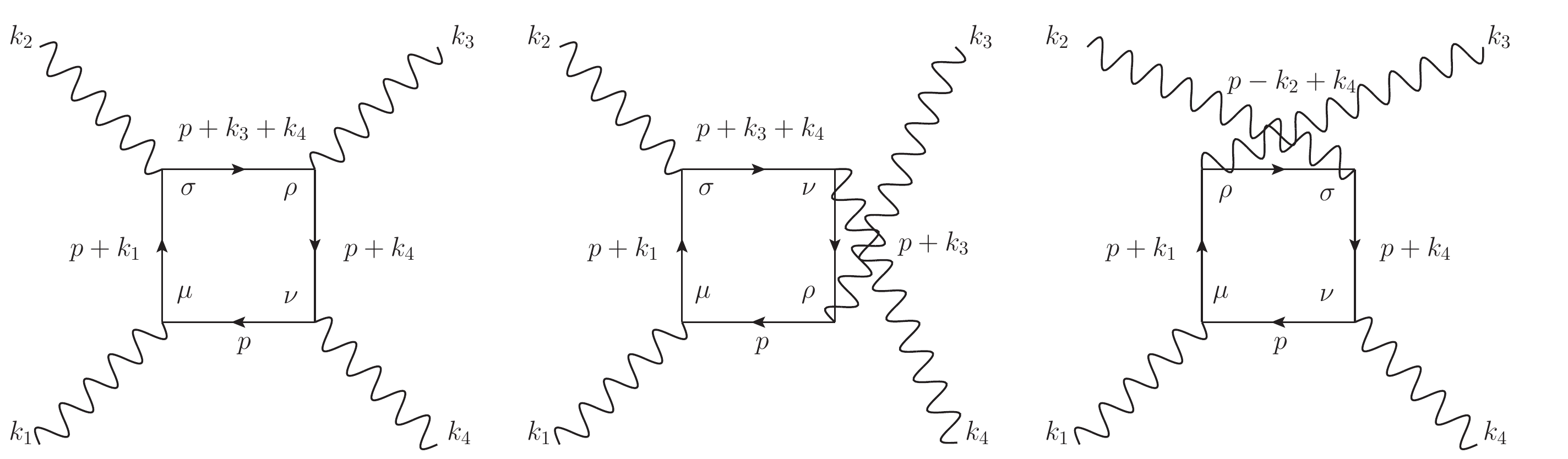}\end{center}
plus the equal charge conjugated versions. The total amplitude is then
\begin{equation}
\mathcal{M}^{\mu\nu\rho\sigma}=2(\mathcal{M}_1^{\mu\nu\rho\sigma}+\mathcal{M}_2^{\mu\rho\nu\sigma}+\mathcal{M}_3^{\mu\nu\sigma\rho}).
\end{equation}
We may calculate the first diagram's amplitude following the standard procedure; we will use similar notation to that of \cite{Kanda,Liang}. We find
\begin{equation}
\begin{split}
&i\mathcal{M}_1^{\mu\nu\rho\sigma}=(iq^{a_1})(iq^{a_2})(iq^{a_3})(iq^{a_4})\\
&\quad\times\int \frac{d^4p}{(2\pi)^4}\frac{\Tr[\gamma^{\mu}(\s
    p+m)\gamma^{\nu}(\s p+\s k_4+m)\gamma^{\rho}(\s p+\s k_3+\s
    k_4+m)\gamma^{\sigma}(\s p+\s
    k_1+m)]}{[p^2-m^2][(p+k_4)^2-m^2][(p+k_3+k_4)^2-m^2][(p+k_1)^2-m^2]}.
\end{split}
\end{equation}
We now introduce Feynman parameters 
to combine the denominators and shift the loop momentum:
\begin{equation}\begin{split}
i\mathcal{M}_1^{\mu\nu\rho\sigma}&=(\prod q^{a_i})3!
\int_0^1 dx_1\int_0^{x_1} dx_2\int_0^{x_2} dx_3 \int \frac{d^4q}{(2\pi)^4}\frac{1}{(q^2-\Delta_1)^4}\\
&\quad\times\Tr[\gamma^{\mu}(\s
    q +\s a_1+m)\gamma^{\nu}(\s q+\s b_1+m)\gamma^{\rho}(\s q+\s
    c_1+m)\gamma^{\sigma}(\s q+\s d_1+m)],\end{split}
\end{equation}
where here
\begin{gather}
A_1=-x_1k_4-x_2k_3+x_3k_2,\\
\Delta_1=m^2-s(x_2-x_3)+x_1x_2s+x_1x_3t+x_2x_3u\equiv m^2\left(1+\frac{U_1}{m^2}\right),\\
a_1=A_1,\quad b_1=A_1+k_4,\quad c_1=A_1+k_3+k_4,\quad d_1=A_1+k_1.
\end{gather}
The traces in the numerator can be simplified by discarding terms odd in the integration variable $q^{\mu}$ as well as odd products of gamma matrices. We may also symmetrize
\begin{equation}
q^{\mu}q^{\nu}\to\frac{1}{4}q^2g^{\mu\nu},\quad
q^{\mu}q^{\nu}q^{\rho}q^{\sigma}\to\frac{1}{24}q^4(g^{\mu\nu}g^{\rho\sigma}+g^{\mu\rho}g^{\nu\sigma}+g^{\mu\sigma}g^{\nu\rho})\end{equation}
to obtain
\begin{equation}\begin{split}
\Tr[\gamma^{\mu}(\s
    q +\s a_1+m)\gamma^{\nu}(\s q+\s b_1+m)&\gamma^{\rho}(\s q+\s
    c_1+m)\gamma^{\sigma}(\s q+\s
    d_1+m)]\\
&=\frac{4}{3}q^4Q^{\mu\nu\rho\sigma}+q^2F_1^{\mu\nu\rho\sigma}+G_1^{\mu\nu\rho\sigma},\end{split}
\end{equation}
where we define the tensors
\begin{align}
F_1^{\mu\nu\rho\sigma}&=-4m^2Q^{\mu\nu\rho\sigma}-\frac{1}{2}R_1^{\mu\nu\rho\sigma},\\
G_1^{\mu\nu\rho\sigma}&=S_1^{\mu\nu\rho\sigma}+m^2P_1^{\mu\nu\rho\sigma}+m^4T^{\mu\nu\rho\sigma},\\
Q^{\mu\nu\rho\sigma}&=g^{\mu\nu}g^{\rho\sigma}+g^{\mu\sigma}g^{\nu\rho}-2g^{\mu\rho}g^{\nu\sigma},\\
\begin{split}
R_1^{\mu\nu\rho\sigma}&=\Tr[\gamma^{\mu}\s
    a_1\gamma^{\nu}(\gamma^{\rho}\gamma^{\sigma}\s d_1+\gamma^{\sigma}\s
      c_1\gamma^{\rho}+\s b_1\gamma^{\rho}\gamma^{\sigma})\\
&\quad+\gamma^{\mu}(\gamma^{\nu}\gamma^{\rho}\s c_1\gamma^{\sigma}\s d_1+\gamma^{\rho}\s b_1\gamma^{\nu}\gamma^{\sigma}\s d_1+\gamma^{\sigma}\s c_1\gamma^{\rho}\s b_1\gamma^{\nu})],\end{split}\\
\begin{split}
P_1^{\mu\nu\rho\sigma}&=\Tr[\gamma^{\mu}\s a_1\gamma^{\nu}(\s
  b_1\gamma^{\rho}\gamma^{\sigma}+\gamma^{\rho}\s
  c_1\gamma^{\sigma}+\gamma^{\rho}\gamma^{\sigma}\s d_1)\\
&\quad+\gamma^{\mu}\gamma^{\nu}\s b_1\gamma^{\rho}(\s c_1\gamma^{\sigma}+\gamma^{\sigma}\s d_1)+\gamma^{\mu}\gamma^{\nu}\gamma^{\rho}\s c_1\gamma^{\sigma}\s d_1],\end{split}\\
S_1^{\mu\nu\rho\sigma}&=\Tr[\gamma^{\mu}\s a_1\gamma^{\nu}\s b_1\gamma^{\rho}\s c_1\gamma^{\sigma}\s d_1],\\
T^{\mu\nu\rho\sigma}&=\Tr[\gamma^{\mu}\gamma^{\nu}\gamma^{\rho}\gamma^{\sigma}]=4(g^{\mu\nu}g^{\rho\sigma}-g^{\mu\rho}g^{\nu\sigma}+g^{\mu\sigma}g^{\nu\rho}).
\end{align}
Evaluating the loop integrals gives
\begin{align}
i\mathcal{M}_1^{\mu\nu\rho\sigma}&=i\frac{\prod q^{a_i}}{(4\pi)^2}\int dS\left[8Q\left(\ln\frac{\Lambda^2}{\Delta}-\frac{11}{6}\right)-\frac{2F_1}{\Delta}+\frac{G_1}{\Delta^2}\right]\\
&=i\frac{\prod q^{a_i}}{(4\pi)^2}\int
dS\left[8Q\left(\ln\frac{\Lambda^2}{m^2}-\ln\left(1+\frac{U_1}{m^2}\right)-\frac{11}{6}\right)\right.\\
&\quad\left.+\frac{8Q+R_1/m^2}{1+U_1/m^2}+\frac{T+P_1/m^2+S_1/m^4}{(1+U_1/m^2)^2}\right].
\end{align}
Despite the superficial logarithmic divergence, the total amplitude is actually finite, since $Q^{\mu\nu\rho\sigma}+Q^{\mu\rho\nu\sigma}+Q^{\mu\nu\sigma\rho}=0$. Nevertheless, contrary to the assertion in \cite{Kanda}, we must properly regularize by subtracting the divergent piece $\mathcal{M}_1^{\mu\nu\rho\sigma}(0,0,0,0)$. 
Once regularized, we may expand to the appropriate order in $1/m$, and combine all three partial amplitudes to get
\begin{align}
\mathcal{M}^{\mu\nu\rho\sigma}&=2\frac{\prod
  q^{a_i}}{(4\pi)^2}\sum_{i=1}^3\int
dS\left[\frac{1}{m^2}(P_i-16Q_iU_i^2+R_i-2T_iU_i)\right.\\
&\quad\left.+\frac{1}{m^4}(12Q_iU_i^2+S_i-R_iU_i-2P_iU_i+3T_iU_i^2)\right].
\end{align}
While the traces and Feynman parameter integrals are too tedious to
compute by hand, Mathematica verifies that the $1/m^2$ term vanishes, leaving 342 terms in the final $1/m^4$ result:
\begin{equation}\label{longans}
\mathcal{M}^{\mu\nu\rho\sigma}=2\frac{\prod q^{a_i}}{(4\pi)^2m^4}\sum_{i=1}^3\int dS\left(12Q_iU_i^2+S_i-R_iU_i-2P_iU_i+3T_iU_i^2\right).
\end{equation}
To find the dyonic amplitudes, we must sum over all combinations of 
\beq
\epsilon_{a_1}^{\mu}(\lambda_1)\epsilon_{a_2}^{\nu}(\lambda_2)\epsilon_{a_3}^{\rho}(\lambda_3)\epsilon_{a_4}^{\sigma}(\lambda_4)\mathcal{M}^{\mu\nu\rho\sigma}~,
\eeq
where the $a_i$ correspond to the charge-space indices in
\eqref{magpols} and contract with the $\prod_iq^{a_i}$ in the
amplitude. 
We also choose the reference vectors for the magnetic polarizations
based on the helicity prescription discussed under \eqref{polspinors}
to obtain Lorentz products of the forms in
\eqref{lorentzprods}. Notice that up to a gauge transformation,
\eqref{magpolhel} implies $\bar{\epsilon}^{\pm}=\mp i\epsilon^{\pm}$,
so that we can see where the appearance of $D_{\pm}$ comes from:
\beq
(\epsilon^{\pm}_{\mu})^aq^a=\epsilon^{\pm}_{\mu}Q+\bar{\epsilon}^{\pm}_{\mu}g/\alpha_d=(Q\mp ig/\alpha_d)\epsilon^{\pm}_{\mu}.
\eeq
Simplification of the kinematics with Mandelstam
invariants results in \eqref{helamps1}.

\end{document}